\documentclass[a4paper,aps,prd,showpacs,nofootinbib,preprintnumbers,amsmath,amssymb,mcite,superscriptaddress,12pt]{revtex4-1}

\usepackage{comment}
\usepackage{soul}
\usepackage[table]{xcolor}
\usepackage{graphicx}
\usepackage{dcolumn}
\usepackage{bm}
\usepackage{esvect}
\usepackage{accents}
\usepackage{slashed}
\usepackage{diagbox}
\usepackage{soul}

\usepackage{hyperref}
\hypersetup{colorlinks=true,linkcolor=magenta,anchorcolor=green,citecolor=cyan,filecolor=black,menucolor=black,urlcolor=brown}
\usepackage{slashed}

\newcommand{\dd}{\mathrm{d}}

\newcommand{\be}{\begin{equation}}
\newcommand{\ee}{\end{equation}}
\newcommand{\ba}{\begin{eqnarray}}
\newcommand{\ea}{\end{eqnarray}}

\newcommand{\Mpl}{M_{\textrm{Pl}}}

\usepackage{braket}

\usepackage{caption}
\usepackage{subcaption}

\begin{document}



\begin{center}
{\huge \bf Constraining the chameleon-photon coupling with atomic spectroscopy}
\end{center} 

\vspace{1truecm}
\thispagestyle{empty}
\centerline{\Large 
Benjamin Elder\footnote{\href{mailto:bcelder@hawaii.edu}{\texttt{bcelder@hawaii.edu}}}
and Jeremy Sakstein\footnote{\href{mailto:sakstein@hawaii.edu}{\texttt{sakstein@hawaii.edu}}}
}

\vspace{.5cm}
 
\centerline{{\it Department of Physics and Astronomy, University of Hawai'i,}}
 \centerline{{\it 2505 Correa Road, Honolulu, HI 96822, USA}} 

\vspace{1cm}

\begin{abstract}
We compute bounds from atomic spectroscopy on chameleon fields that couple to the photon.~Chameleons are a wide class of scalar field models that generically lead to {screened} fifth forces and a host of novel phenomenologies, particularly when the photon coupling is included.~We account for perturbations to the atomic energy levels from {both} the scalar field ``fifth force'' and the scalar field's correction to the electric field.~We also account for the electromagnetic interaction's contribution to the scalar charge of the proton, which enables a considerably wider class of models to be tested than without this effect.~We find bounds that cover different areas of chameleon parameter space.~Some regions are redundant with existing experiments, particularly $g - 2$, confirming that those models are ruled out.~Other regions were previously unconstrained, and a range of models spanning approximately four orders of magnitude in chameleon coupling parameters are {excluded} for the first time.
\end{abstract}

\maketitle

\section{Introduction}

Scalar fields are commonly included in new theories of gravity, dark energy, and dark matter.~They represent a minimal modification (i.e.,~a single degree of freedom) to General Relativity and the Standard Model.~It is therefore of paramount importance to have a good understanding of what types of scalar fields can exist in nature and still be compatible with the myriad experimental bounds on new physics.~For example, the principles of effective field theory (EFT) teach us that when writing down the Lagrangians for new physics, we must include all possible couplings that are compatible with the symmetries of the theory.~For a scalar field, absent some special symmetry we must include an explicit coupling between the scalar field $\phi$ and (fermionic) matter fields $\psi$, the {lowest-order} being {a Yukawa interaction}
\begin{equation}
  {\cal L_\mathrm{int}} \supset \frac{m_\psi}{M} \phi \bar \psi \psi,
  \label{scalar-matter-interaction}
\end{equation}
where $m_\psi$ is the mass of the matter field and $M$ is a parameter controlling the coupling strength.~This term implies that the scalar field mediates a ``fifth force'' between matter fields, with a range comparable to the Compton wavelength of the scalar particle.~Experimental searches for fifth forces then imply strong bounds on the mass of the scalar particle and its coupling strength to matter fields~\cite{Adelberger:2003zx, Kapner:2006si, CANTATA:2021ktz}.

The story becomes more complicated when there are other interaction terms in the Lagrangian.~The paradigm of ``screened'' theories~\cite{Joyce:2014kja} has shown that relatively simple additional interactions can render the fifth force difficult to observe in nature, demonstrating that there remain wide classes of light scalar field theories that are compatible with existing experimental tests.~There are three canonical models of screening: (i) the chameleon~\cite{Khoury:2003aq}, in which the scalar particle's mass can vary, making the fifth force short-ranged in certain environments; (ii) the symmetron~\cite{Hinterbichler:2010es}, in which the scalar decouples from matter, making the fifth force weak; and (iii) the galileon~\cite{Nicolis:2008in}, which suppresses gradients in the scalar field.~All of these mechanisms serve to weaken the force in everyday environments, showing that a very wide class of models are compatible with traditional fifth force tests.~In response to this, a great deal of effort on both the experimental and theory sides has gone into detecting the subtle signatures of screened theories.~For recent reviews, see~\cite{Joyce:2014kja, Burrage:2016bwy, Burrage:2017qrf,Sakstein:2018fwz,Baker:2019gxo, Brax:2021wcv, CANTATA:2021ktz}.

It is important to bear in mind that the interaction in Eq.~\eqref{scalar-matter-interaction} is not the only new interaction with Standard Model fields that can be considered.~One could just as easily include a coupling to the photon, with the leading-order term in the Lagrangian being
\begin{equation}
\label{eq:scalar-photon-coupling}
  {\cal L}_\mathrm{int} \supset \frac{\phi}{M_\gamma} F^2,
\end{equation}
where $F$ is the photon's field strength tensor and $M_\gamma$ controls the coupling strength.~In fact, the inclusion of this term opens up a wide range of new and interesting phenomenologies for the theory, as well as a number of new ways to detect the scalar particle.~{For example, Ref.~\cite{Vagnozzi:2021quy} has recently shown that chameleons emitted in the solar tachocline can be detected directly using planned electron-recoil dark matter direct detection chambers e.g., XENONnT.} 

This paper focuses on the scenario in which the new particle couples both to matter fields and to the photon, and includes a screening mechanism.~Experimental constraints on this specific scenario were reviewed in~\cite{Burrage:2017qrf}.~For concreteness, we focus on the chameleon mechanism, although our formalism could be straightforwardly applied to the symmetron as well.\footnote{The same could not be said of the galileon, as its reliance on derivative operators makes the method of analysis used here inapplicable to that case. Furthermore, although it is commonly included the operator in Eq.~\eqref{eq:scalar-photon-coupling} is not compatible with the galileon's symmetry so is absent from the EFT.} This scenario has been tested in a variety of regimes, most significantly (for the chameleon) by the CAST~\cite{CAST:2015npk}, GammeV~\cite{Steffen:2010ze}, and electron $g - 2$ experiments~\cite{Brax:2018zfb}\footnote{There are also older bounds coming from e.g.~the PVLAS experiment~\cite{PVLAS:2005sku,Brax:2007ak,Brax:2007hi}, but the area of parameter space that this experiment tests has been superseded by more recent atom interferometry experiments~\cite{Jaffe:2016fsh,Sabulsky:2018jma}.}.~These are complemented by a host of astrophysical tests as well.~These tests work on different length and energy scales, resulting in mostly complementary constraints in chameleon parameter space.

This paper is concerned with constraining the scalar-photon interaction using atomic spectroscopy, particularly the hydrogen 1s - 2s transition energy, which has been measured to an accuracy of a few parts in $10^{15}$~\cite{Parthey_2011PhRvL.107t3001P}.~To this end, we solve for both the chameleon field around the proton, as well as the modification to the electrostatic potential sourced by the proton.~Both of these then lead to perturbations in the hydrogen spectrum, which we then use to place bounds on the theory.~We find bounds that cover a wide regime of chameleon parameter space.~Some parts are redundant with existing experiments (particularly $g - 2$), which confirms that those models are now conclusively ruled out by two independent experiments.~Meanwhile, other regions of parameter space are constrained for the first time.~It should be noted that Ref.~\cite{Brax:2010gp} also examined atomic spectroscopy in the context of this chameleon model.~The key {novelty of} the present work is that we account for the change to the scalar charge of the proton due to the photon coupling, which leads to stronger constraints.~Atomic spectroscopy bounds on chameleons were also recently considered in~\cite{Brax:2022olf}, but tested a model that did not include a scalar-photon coupling.

This paper is organized as follows.~In Section~\ref{sec:scalar-photon-EM} we briefly compute the perturbations to the electron's Hamiltonian for a generic scalar field.~In Section~\ref{sec:scalar-solution} we focus on the scalar field's profile, allowing for scalar particles that admit a screening mechanism.~In Section~\ref{sec:atomic-perturbations} we compute the perturbation to the $1s$ and $2s$ energy levels by a screened scalar.~In Section~\ref{sec:experimental-constraints} we specialize to a particular class of chameleon models, as well as a particular atomic spectroscopy experiment to produce bounds on that theory.~Those bounds are then compared to other experimental bounds in Sec.~\ref{sec:comparison-to-other-bounds}.~We conclude in Section~\ref{sec:conclusions}.

\section{Scalar and Photon Electrodynamics}
\label{sec:scalar-photon-EM}
We begin with a generic a scalar field theory that is defined by the action\footnote{We work in the mostly-positive metric convention, as well as with units where $\hbar = c = 1$.~We also define the reduced Planck mass as $\Mpl = (8 \pi G)^{-1/2}$.}
\begin{equation}
  S = \int d^4 x \sqrt{-g} \left( -\frac{1}{4}F^2 + A_\mu J^\mu - \frac{1}{2} (\partial \phi)^2 - V(\phi) - \frac{\rho}{M} \phi - \frac{\phi}{4 M_\gamma} F^2 \right).
\end{equation} 
We will {now} solve for the scalar field around a spherical charge of constant density {(our model for the proton)}, and also for the correction to the electromagnetic field $F_{\mu \nu}$.
The equations of motion are
\begin{align} \nonumber
  J^\nu &= \partial_\mu \left( \left(1 + \frac{ \phi}{M_\gamma} \right) F^{\nu \mu} \right), \\
  \Box \phi &= V,_\phi + \frac{\rho}{M} + \frac{1}{4 M_\gamma} F^2.
\end{align}
The vector EOM contains the equation
\begin{equation}
  \rho_q = \vec \nabla \cdot \left( \left(1 + \frac{\phi}{M_\gamma} \right) \vec E \right),
\end{equation}
where $\rho_q$ is the electric charge.~A Gaussian sphere enclosing a charge $Q_\mathrm{encl}$ has an electric field of magnitude
\begin{equation}
  E = \frac{Q_\mathrm{encl}}{4 \pi r^2} \frac{1}{1 + \phi / M_\gamma} \approx \frac{Q_\mathrm{encl}}{4 \pi r^2} \left(1 - \frac{\phi}{M_\gamma} \right),
\end{equation}
normal to {its surface.} We define the {unperturbed} electric field $\bar E$ and its perturbation $\delta E$ as 
\begin{align}\nonumber
  E &= \bar E + \delta E, \\ 
  \bar E &= \frac{Q_\mathrm{encl}}{4 \pi r^2}, \quad
  \delta E = - \bar E \frac{\phi}{M_\gamma}.
  \label{E-definition}
\end{align}
We also define the electrostatic potential in the usual way
\begin{equation}
  V(r) = - \int_{\cal \infty}^r \vec E \cdot \vec{\dd l},
\end{equation}
so that we have $V = 0$ at spatial infinity.~We will again split {this} up as $V = \bar V + \delta V$, and using Eq.~\eqref{E-definition} we have
\begin{align} \nonumber
  V &= -\frac{Q}{4 \pi r} + \delta V, \\
  \delta V &\equiv \frac{1}{4 \pi M_\gamma} \int_\infty^r \frac{ \phi Q_\mathrm{encl}}{r'^2} \dd r'.
  \label{V-and-perturbation}
\end{align}

In this work we will be concerned with hydrogen, where a single electron orbits a single proton.~The proton is modelled as a sphere of radius $R$ with uniform mass and charge densities, for a total mass $m_p$ and {total} charge $Q$.~At leading order, the electric field is
\begin{equation}
  \bar E = \begin{cases}
  \frac{Q r}{4 \pi R^3} & r \leq R, \\
  \frac{Q}{4 \pi r^2} & r > R.
  \end{cases}
  \label{E-field-zeroth}
\end{equation}

Our aim is to perform non-relativistic perturbation theory on the electron with mass $m_e$ and charge $q$, which has Hamiltonian 
\begin{equation}
  H = \frac{p^2}{2 m_e} + \frac{ m_e }{M} \phi + q V.
\end{equation}
Using Eq.~\eqref{V-and-perturbation}, we identify two perturbations to the Hamiltonian:
\begin{equation}
  \delta H_m = \frac{m_e}{M} \phi, \quad \quad \delta H_\gamma = q \delta V.
\end{equation}

\section{Scalar field solution}
\label{sec:scalar-solution}

To proceed we must provide a particular solution of the scalar field.~To deal with the field's potentially nonlinear equation of motion, we solve in a piecewise manner inside and outside the proton, where in each region we have linearized the theory about some scalar field value:
\begin{equation}
  \phi = \bar \phi + \varphi.
\end{equation}
The external field profile is simplest to deal with as $\rho = 0$ there.~We linearize about the ambient scalar field value $\bar \phi_\mathrm{out}$ far away from the atom.~Furthermore, we assume that the scalar field's Compton wavelength is much larger than the size of the atom, that is, $m_\mathrm{out}^{-1} \gg a_0$, where $m$ is the effective mass of the scalar field, $m^2(\bar \phi) = \frac{d^2}{d \phi^2} V(\phi)\big|_{\phi = \bar \phi}$ and $a_0$ is the Bohr radius.~It follows that the external field equation of motion is 
\begin{equation}
  \Box \varphi = - \frac{1}{2 M_\gamma} \bar E^2,
\end{equation}
where we have used $F^2 = 2 (\vec B^2 - \vec E^2)$.~The solution is
\begin{equation}
  \phi(r > R) = \bar \phi_\mathrm{out} - \frac{B}{r} - \frac{Q^2}{64 \pi^2 M_\gamma r^2}.
\end{equation}
The integration constant $B$ is the monopole of the proton's field configuration, and will be solved for by matching to the interior solution.

There are two cases to consider for the interior solution.~In the first case, the scalar field's Compton wavelength inside the proton is smaller than the proton radius.~In this regime the field rolls to its equilibrium value in the central region of the proton so we expand about the equilibrium field value $\bar \phi_\mathrm{in}$, which is defined via
\begin{equation}
   \frac{d}{d \phi} V_\mathrm{eff}(\bar \phi_\mathrm{in})=0,
\end{equation}
where $V_\mathrm{eff}(\phi) = V(\phi) + \frac{\phi}{M} \rho$.~Note that we are ignoring the $F^2$ term in the effective potential, which is only appropriate if the electromagnetic coupling is sufficiently weak.~Using Eq.~\eqref{E-field-zeroth} for the magnitude of the electric field, this condition is satisfied everywhere within the proton provided that $M_\gamma / M \gg E^2 / \rho \approx 10^{-4}$, which corresponds to a region of parameter space that is largely complementary to existing constraints from other experiments.~Having linearized about $\bar \phi_\mathrm{in}$, the equation of motion is
\begin{equation}
  \Box \varphi = m^2_\mathrm{in} \varphi - \frac{1}{2 M_\gamma} \bar E^2.~
\end{equation}
The solution to this equation is
\begin{equation}
  \phi(r < R) = \bar \phi_\mathrm{in} - \frac{A \sinh m_\mathrm{in} r}{r} + \frac{Q^2 \left(6 + m_\mathrm{in}^2 r^2 \right)}{32 \pi^2 M_\gamma m_\mathrm{in}^4 R^6},
\end{equation}
where {$A$ is an integration constant and} we have enforced the condition $\phi'(0) = 0$. {Matching} $\phi$ and $\phi'$ at the boundary $r = R$, we find, in the limits $m_\mathrm{in} R \gg 1$ and $|\bar \phi_\mathrm{in}| \ll |\bar \phi_\mathrm{out}|$, a monopole
\begin{equation}
  B_\mathrm{screened} = \bar \phi_\mathrm{out} R - \frac{Q^2}{64 \pi^2 M_\gamma R}.
\end{equation}
The first term is the standard monopole term for a strongly-screened scalar field.~The second term, however, is new and depends solely on the photon coupling $M_\gamma$.

We now consider the other limit, in which $m_\mathrm{in} R \ll 1$, such that the scalar field does not deviate {significantly} from the ambient field value.~In this case, we expand about the external field value $\bar \phi_\mathrm{out}$.~This is not an equilibrium field value, and consequently the equation of motion for the scalar field in the interior region is
\begin{equation}
  \Box \varphi = J - \frac{2}{M_\gamma} \bar E^2,
\end{equation}
where we have defined $J \equiv \frac{d}{d \phi} V_\mathrm{eff}(\bar \phi_\mathrm{out})$.~In this case the interior scalar field is
\begin{equation}
  \phi(r < R) = \bar \phi_\mathrm{out} + A + \frac{Jr^2}{6} - \frac{Q^2 r^4}{640 \pi^2 M_\gamma R^6}.
\end{equation}
Once again $A$ is an integration constant, and we have enforced the condition $\phi'(0) = 0$.~The same procedure of matching $\phi$ and its first derivative at $R$ gives a monopole
\begin{equation}
  B_\mathrm{unscreened} = \frac{R^3 J}{3} - \frac{3 Q^2}{80 \pi^2 M_\gamma R}.
\end{equation}
The quantity $J$ is dominated by the density of the object $\rho$, so the first term represents the usual monopole for an unscreened scalar field coupled to matter, $m_\mathrm{p} / M$.~We find once again a monopole sourced by the electric field around the proton.

We can cover both of these regimes by the single expression
\begin{equation}
  \phi(r > R) = \bar \phi_\mathrm{out} - \frac{B}{r} - \frac{\alpha}{16 \pi M_\gamma r^2},
  \label{phi-exterior-soln}
\end{equation}
where we have used $Q = e$ and the fine-structure constant $\alpha = \frac{e^2}{4 \pi}$ to simplify the last term.~This covers both regimes by writing the monopole as
\begin{equation}
  B = \frac{1}{4 \pi} \left( \frac{\lambda m_p}{M} - f(\lambda) \frac{\alpha}{M_\gamma R} \right),
\end{equation}
where we have introduced the {\it screening factor} of the proton
\begin{equation}
  \lambda = \min \left( 4 \pi \bar \phi_\mathrm{out} R \frac{M}{m_p}, 1 \right),
\end{equation}
and have introduced the factor
\begin{equation}
  f(\lambda) = \begin{cases}
  1/4 & \lambda < 1 , \\
  3/5 & \lambda = 1 .
  \end{cases}
\end{equation}
In both of these regimes we have found that there is a term in the monopole that scales as $\alpha / (M_\gamma R)$.
This term decays more slowly than the $1/r^2$ term in Eq.~\eqref{phi-exterior-soln} and is therefore the dominant contribution from the photon coupling to the scalar field configuration.~This is the key result upon which we will derive experimental constraints.

\section{Perturbation to atomic energy levels}
\label{sec:atomic-perturbations}
Our aim is to do perturbation theory on the non-relativistic electron with mass $m$ and charge $q$ (for an electron, $q = -e$). A non-relativistic particle has a Hamiltonian
\begin{equation}
  H = \frac{p^2}{2 m} + \frac{ m_e }{M} \phi + q V .
\end{equation}
We see that the scalar field {contributes two perturbations to} the Hamiltonian. The first is via the explicit matter coupling:
\begin{equation}
  \delta H_m = \frac{m_e}{M} \phi = \frac{m_e B}{M} \frac{1}{r}.
\end{equation}
The second results from perturbing the electrostatic potential $V$, which in turn perturbs the Hamiltonian by an amount
\begin{align}
  \delta H_\gamma = q \delta V = \frac{\alpha B}{M_\gamma} \frac{1}{r^2}.
\end{align}
Note that when substituting in the solution for $\phi$ we have {neglected} the irrelevant constant term $\bar \phi_\mathrm{out}$, as well as the sub-leading $1/r^2$ term.~This  {approximation} is {justified} via explicit calculation in Appendix~\ref{appendix-finite-nucleus}.

The perturbations to the electron's energy levels are computed via
\begin{equation}
  \delta E_n = \bra{\psi_n} \delta H \ket{\psi_n}.
\end{equation}
The tightest experimental bounds from spectroscopy are on the hydrogen $1s$-$2s$ energy levels, which have wavefunctions given by\footnote{Earlier we saw that including the finite size of the nucleus had important consequences for the scalar charge of the object.~The finite size of the nucleus would also slightly modify these wavefunctions.~Including those modifications here would shift our results by at most 1 part in $10^5$, so for the present purposes we stick to wavefunctions around an infinitely small proton.~This is shown explicitly in Appendix~\ref{appendix-finite-nucleus}.}
\begin{align} \nonumber
  \psi_{1s} &= \frac{1}{\sqrt \pi a_0^{3/2}} e^{- r / a_0} ,\\
  \psi_{2s} &= \frac{1}{4 \sqrt{2 \pi} a_0^{3/2}} \left(2 - \frac{r}{a_0} \right) e^{- r / (2 a_0)}.
\end{align}
The energy levels are shifted by an amount
\begin{align} \nonumber
  \delta E_1 &= \frac{m_e B}{M} \frac{1}{a_0} + \frac{\alpha B}{M_\gamma} \frac{2}{a_0^2} ,\\
  \delta E_2 &= \frac{m_e B}{M} \frac{1}{4 a_0} + \frac{\alpha B}{M_\gamma} \frac{4}{a_0^2},
  \label{energy-level-shifts}
\end{align}
where we have used the charges of the electron and proton ($q = -e, Q = +e$ respectively), and have also made use of the integrals given in Table~\ref{tab:integrals}.~The perturbation to the transition energy between the $1s$ and $2s$ orbitals is
\begin{align} \nonumber
  \delta E_{1-2} &= |\delta E_2 - \delta E_1| = \frac{m_e B}{M} \frac{3}{4a_0} + \frac{\alpha B}{M_\gamma} \frac{2}{a_0^2},\\
  &= \frac{3 m_e \lambda m_p}{16 \pi M^2 a_0} - \frac{3 f(\lambda) m_e \alpha}{16 \pi M M_\gamma a_0 R} + \frac{ \lambda m_p \alpha}{2 \pi M M_\gamma a_0^2} - \frac{f(\lambda) \alpha^2}{2 \pi M_\gamma^2 a_0^2 R}.
  \label{deltaE-final}
\end{align}
The first term is purely a result of the matter coupling, and has been explored in detail before~\cite{Brax:2010gp, Wong:2017jer, Brax:2022olf}.~Our interest is the photon coupling $M_\gamma$, so we will focus on the remaining three terms.~The third term has {been studied previously}, but we can see that it is always smaller than the second term, as $f(\lambda) \approx 1$ and $\lambda$ is at most 1.~The second and fourth terms are new and will result in new bounds on the photon coupling.

\begin{table}[]
  \centering
  \begin{tabular}{| c || c c |}
\hline
      & $m = 1$ & $m = 2$ \\ 
      \hline
 $n = 1$ & $\frac{1}{a_0}$ & $\frac{2}{a_0^2}$ \\ 
 $n = 2$ & $\frac{1}{4 a_0}$ & $\frac{4}{a_0^2}$ \\
 \hline
  \end{tabular}
  \caption{Solutions to the integral $\bra{\psi_n} r^{-m} \ket{\psi_n}$ for the $s$ orbitals of the hydrogen atom.}
  \label{tab:integrals}
\end{table}

\section{Experimental Constraints}
\label{sec:experimental-constraints}

The hydrogen $1s$-$2s$ energy level difference has been experimentally measured to a relative accuracy of $4.2 \times 10^{-15}$~\cite{Parthey_2011PhRvL.107t3001P}.~This can be combined with Eq.~\eqref{deltaE-final} to constrain screened theories.

In order to proceed, we must specify a scalar potential.~For concreteness, we will focus on the typical chameleon model:
\begin{equation}
  V(\phi) = \frac{\Lambda^5}{\phi}.
\end{equation}
This enables us to determine the ambient scalar field value $\phi_\mathrm{out}$ far from the hydrogen atom.

The chameleon potential is shaped such that the field rolls towards the minimum of its ``effective potential'' which is the sum of its self-coupling potential $V(\phi)$ and its couplings to other fields, such as to matter and to photons.~In the present work we focus on the regime in which the coupling to matter dominates the coupling to photons, so we have
\begin{equation}
  V_\mathrm{eff}(\phi) = \frac{\Lambda^5}{\phi} + \frac{\phi}{M} \rho.
\end{equation}
In a region of constant matter density $\rho$, such as a vacuum chamber where the hydrogen spectroscopy measurement is performed, the field rolls to a value that minimises its potential
\begin{equation}
  \phi_\mathrm{min} = \left( \frac{M \Lambda^5}{\rho} \right)^{1/2}.
\end{equation}
Scalar perturbations around any particular field value have a mass
\begin{equation}
  m_\mathrm{eff}(\phi) = \left( \frac{2 \Lambda^5}{\phi^3} \right)^{1/2}.
\end{equation}
However, inside a given vacuum chamber, the field may not have sufficient room to reach $\phi_\mathrm{min}$.~The field profile is one that minimizes its Hamiltonian, which means balancing gradient energy ($|\vec \nabla \phi|^2$) against its potential energy ($V_\mathrm{eff}$).~If the field's Compton wavelength about $\phi_\mathrm{min}$ is larger than the vacuum chamber's characteristic inner dimension $R_\mathrm{vac}$, then the lowest energy configuration available to the chameleon field is to roll to a value $\phi_\mathrm{vac}$ such that $m_\mathrm{eff}(\phi_\mathrm{vac})^{-1} \approx R_\mathrm{vac}$~\cite{Hamilton:2015zga, Elder:2016yxm}.~Explicitly, we have
\begin{equation}
  \phi_\mathrm{vac} = \xi (2 \Lambda^5 R_\mathrm{vac}^2)^{1/3},
\end{equation}
for an $O(1)$ constant $\xi$ which depends on the geometry of the vacuum chamber.~We approximate the vacuum chamber to be spherical, so $\xi = 0.55$ \cite{Hamilton:2015zga,Elder:2016yxm}. The vacuum chamber and internal apparatus of~\cite{Parthey_2011PhRvL.107t3001P} is not spherical, so we set$R_\mathrm{vac} = 1~\mathrm{mm}$ which is approximately the distance between the atoms and other parts of the internal measurement apparatus~\cite{Parthey:2010aya}.  The ambient field value in the center of the vacuum chamber will be whichever of these two field values is smaller:
\begin{equation}
  \phi_\mathrm{amb} = \min( \phi_\mathrm{min}, \phi_\mathrm{vac} ).
\end{equation}
This is the field value we use when computing the screening factor $\lambda$ in Eq.~\eqref{deltaE-final} and comparing against the experimental uncertainty of this quantity, stated at the beginning of this section.

\begin{figure}[t]
\centering
\begin{subfigure}{0.45\textwidth}
  \centering
  \includegraphics[height=2.75in]{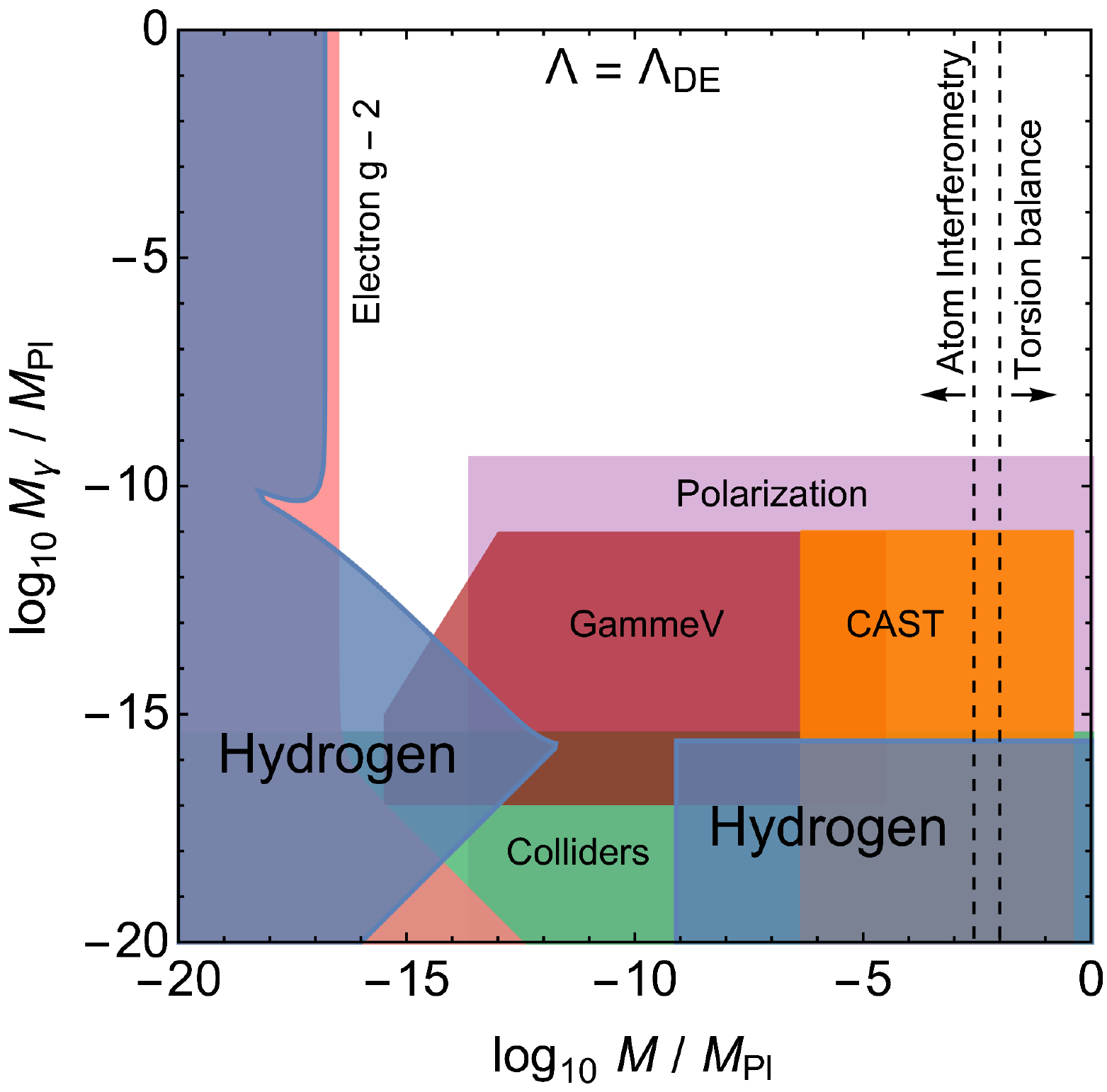}
  \caption{}
  \label{fig:constraints-meV}
\end{subfigure}
\begin{subfigure}{0.45\textwidth}
  \centering
  \includegraphics[height=2.75in]{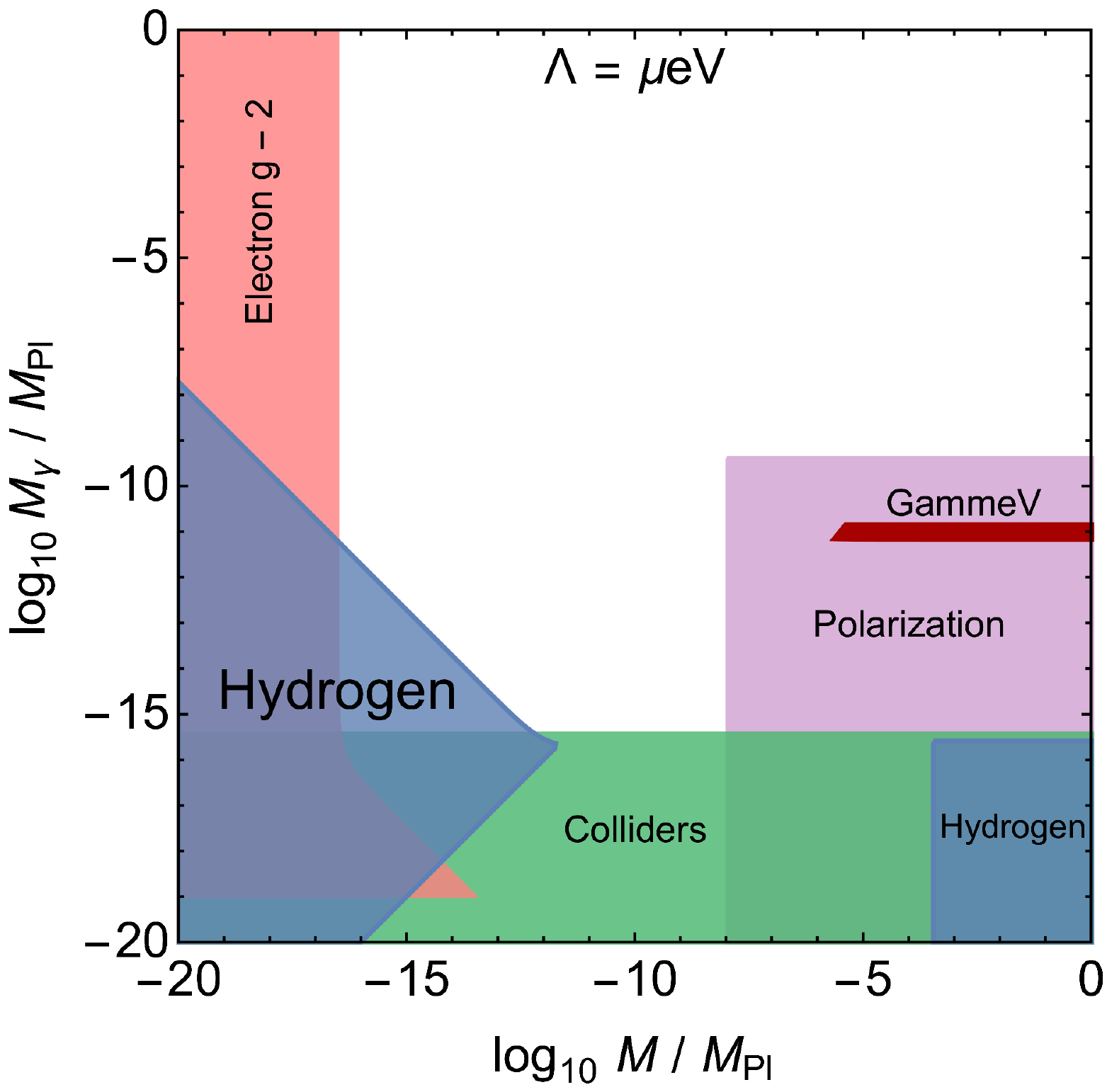}
  \caption{}
  \label{fig:constraints-microeV}
\end{subfigure}
\caption{\small Combined constraints on chameleon parameter space, with the chameleon self-coupling parameter $\Lambda$ set to the dark energy scale $\Lambda_\mathrm{DE} = 2.4~\mathrm{meV}$ (left) and to $\Lambda = \mu{\mathrm{eV}}$ (right). }
\label{fig:constraints}
\end{figure}

The resulting constraint on the chameleon's parameter space are plotted in Fig.~\ref{fig:constraints}.~We have focused on two particular choices for the chameleon self-coupling parameter, $\Lambda = 2.4~\mathrm{meV}$ and $\Lambda = \mu \mathrm{eV}$.~The first is a standard choice, corresponding to the dark energy scale, and has been extensively tested.~Fifth-force searches have placed powerful bounds that are independent of the photon coupling $M_\gamma$.~Atom interferometry and torsion balance experiments are highly complementary and leave a small window of unconstrained parameter space between them.~This window has recently been tested using a levitated force sensor~\cite{Yin:2022geb}, and may be further tested in the near future with Casimir force sensors~\cite{Brax:2022uiv}. The second choice, $\Lambda = \mu \mathrm{eV}$, corresponds to the region of parameter space where solar chameleons could be detected using planned electron-recoil dark matter direct detection chambers \cite{Vagnozzi:2021quy}. These plots show that hydrogen spectroscopy is complementary to existing tests in certain regions.~In other regions, it re-tests some areas that have been covered by previous experiments, like electron $g - 2$, GammeV, and CAST.

\section{Comparison to other experimental bounds}
\label{sec:comparison-to-other-bounds}

It will be useful to compare our constraints to existing bounds.~However, these bounds are often not presented in full generality, either stating their bounds in terms of the local chameleon mass or by fixing one of the chameleon parameters (typically by setting $\Lambda = 2.4~\mathrm{meV}$).~As we have adopted neither of these approaches, it will be necessary to reinterpret some of these constraints.  A reader who is already familiar with these experiments and their constraints may wish to skip ahead to our conclusions in Sec.~\ref{sec:conclusions}.

\subsection{GammeV-CHASE}
The GammeV-CHASE experiment uses a laser and a magnetic field to produce chameleon particles inside a vacuum chamber.~After a period of time, the laser is switched off and the conversion of chameleons back into photons may be detected.~This experiment is sensitive to models where (i) the chameleon mass in the vacuum chamber is sufficiently small to be produced by the laser, and (ii) the chameleon mass inside the vacuum chamber walls is sufficiently large to trap the chameleons inside the chamber.~Furthermore, it required that (iii) the chameleon-photon coupling not be too large, otherwise the conversion of chameleons into photons occurs too quickly to be observed.~The constraints in Fig.~\ref{fig:constraints-meV} are taken directly from the most recent publication~\cite{Steffen:2010ze}.~This paper did not report constraints at the $\Lambda = \mu \mathrm{eV}$ scale, although some constraints are expected to still apply within the appropriate regime of parameter space in which the above-mentioned criteria are satisfied.~The constrained region in Fig.~\ref{fig:constraints-microeV} is the region in which conditions (i) - (iii) are satisfied, and where the photon coupling is $M_\gamma > 2 \times 10^{-12} \Mpl$~\cite{GammeV:2008cqp}.~A more detailed analysis, of the sort done in \cite{Steffen:2010ze}, would likely exclude a slightly larger region of parameter space, but is beyond the scope of the present work.

\subsection{CAST}
The CERN Axion Solar Telescope (CAST) is sensitive to chameleons produced in the Sun.~Strong electromagnetic fields in the Sun produce chameleons via the electromagnetic coupling, which then propagate inside the CAST detector, which has a strong magnetic field to convert the chameleon particles back into photons which are then measured by the detector.~Like the GammeV experiment, this relies on a balance of different behaviors: the chameleons must be too heavy to be produced in the core of the Sun (otherwise they would cause stars to cool too quickly) yet also be light enough to be produced in the Sun's tachocline, and sufficiently weakly coupled to matter to be able to propagate through the lead shielding of the detector.~The latest constraint on chameleons~\cite{CAST:2015npk} is shown in Fig.~\ref{fig:constraints-meV}.~That paper did not provide constraints for the $\Lambda = \mu\mathrm{eV}$ model, although to be consistent with the above requirements the constraint would exist within 1-2 orders of magnitude in $M$ around $M \approx \Mpl$.~

\subsection{Particle Colliders}
Very similar models to the one considered here have been constrained using data from {particle} colliders~\cite{Brax:2009aw}.~This analysis ruled out models with $M_\gamma \gtrsim ~\mathrm{TeV}$.~However, there are two important caveats to note.~First, the chameleon model was assumed to have a mass smaller than 
$10^{-8}~\mathrm{eV}$ inside the collider ring.~The models considered here {are} considerably heavier in that environment, and therefore these bounds relax by an $O(1)$ factor~\cite{Brax:2009aw}.~Second, that study relied on a uniform coupling to all Standard Model gauge fields, not just the photon as we consider here.~As such, this bound does not conclusively rule out any of this parameter space.

\subsection{Electron $g - 2$}
Measurements of the anomalous magnetic moment of the electron are capable of constraining the chameleon's couplings to matter and to photons~\cite{Brax:2018zfb}.~The chameleon contributes both a quantum and classical effect to the measurement, although for the two values of $\Lambda$ chosen here only the quantum contribution is relevant.~When the chameleon mass is much smaller than the electron mass, the contribution is
\begin{equation}
  \delta a_\mu = 3 \left( \frac{m_e}{4 \pi M} \right)^2 + 4 \frac{1}{M_\gamma M} \left(\frac{m_e}{4 \pi} \right)^2 \left( O(1) + \frac{3}{2} \right).
\end{equation}
The experimental constraint on the anomalous magnetic moment is $\delta a_\mu < 0.77 \times 10^{-12}$~\cite{2008PhRvL.100l0801H, Hanneke:2010au}, yielding the constraints shown in Fig.~\ref{fig:constraints}.

\subsection{Polarization}
The chameleon-photon interaction can also result in the polarization of light from distant stars~\cite{Burrage:2008ii}.~It has been shown that this results in a constraint that rules out $M_\gamma < 1.1 \times 10^{9}~\mathrm{GeV}$ for chameleon models that have an effective mass $m_\phi < 1.3 \times 10^{-11}~\mathrm{eV}$ inside the Milky Way~\cite{Burrage:2008ii,Burrage:2017qrf}, where the ambient density was taken to be $10^{-24}~\mathrm{g/cm}^3$.~This constraint, along with constraints coming from measurements of active galactic nuclei~\cite{Burrage:2009mj,Pettinari:2010ay} involve assumptions about generally poorly-known variables like the ambient densities and magnetic fields in astrophysical environments, so although the regions are suggested to be excluded one cannot say that they are conclusively ruled out~\cite{Burrage:2017qrf}.

\section{Discussion and Conclusions}
\label{sec:conclusions}

We have constrained a wide range of chameleon models that couple to photons with existing experimental bounds on the hydrogen spectrum.~This was {achieved} by solving for the chameleon field around a hydrogen nucleus, as well as the shifts to the electron energy levels.~There are several effects that produce perturbations:~first, the scalar coupling to matter, which mediates a fifth force between the proton and electron, as well as a shift to the electrostatic potential due to the presence of the scalar field profile.~We focused on the case in which the scalar field profile is determined by its matter coupling, and thus there is the usual chameleon phenomenology in which the proton can be screened or unscreened depending on a model's specific values of the self-coupling parameter $\Lambda$ and the matter coupling $M_\gamma$.~In future work it would be interesting to relax this assumption, which would widen the scope of constrainable parameter space and allow the intriguing possibility for the photon interactions to screen (or unscreen) the proton, and charged particles more generally.~This work also showed the possibility for the chameleon to behave like a varying-$\alpha$ theory, although it was not investigated.~A thorough investigation of this intriguing phenomenon would be of great interest.

We have placed new bounds on chameleon parameter space, {and} confirmed that existing models previously tested by electron $g - 2$ experiments, GammeV, and CAST are now conclusively ruled out.~At the dark energy scale, $\Lambda = 2.4~\mathrm{meV}$, there are now a great deal of experimental tests that put the simplest chameleon models considered here under significant tension.~We have shown how our new bounds change with $\Lambda$ by also presenting the parameter space at $\Lambda = \mu \mathrm{eV}$, {which may also be probed with planned dark matter direct detection experiments}.~Where possible, we have included bounds from other experiments, but were forced to make estimates of certain other experiments' capabilities in this regime, particularly GammeV and CAST.~It would be interesting to see a detailed analysis of these experiments in this region of parameter space, and how they compare to the bounds presented here.

{\small {\bf Acknowledgments} The authors are grateful to Xerxes Tata and Clare Burrage for helpful discussions.}

\appendix
\section{Scalar field solution -- finite sized nucleus, screened case}
\label{appendix-finite-nucleus}
When we computed the scalar charge of the proton, we found that it was important to account for its finite size.~However, when we computed the perturbations to the energy levels, we treated the exterior solution as valid all the way down to $r = 0$.~We also neglected the sub-leading $r^{-2}$ term in the exterior solution when computing these integrals.~In this Appendix we show that both of these approximations were justified.

We begin with the screened case, where the full solution is
\begin{equation}
  \phi(r) = \begin{cases}
  \bar \phi_\mathrm{in} - \frac{A \sinh m_\mathrm{in} r}{m_\mathrm{in} r} + \frac{Q^2 (6 + m_\mathrm{in}^2 r^2)}{32 \pi^2 M_\gamma m_\mathrm{in}^4 R^6} & r \leq R, \\
  \bar \phi_\mathrm{out} - \frac{B}{r} - \frac{Q^2}{64 \pi^2 M_\gamma r^2} & r > R .
  \end{cases}
\end{equation}
The integration constants are
\begin{align} \nonumber
  A &= \frac{2 m_\mathrm{in}}{m_\mathrm{in} R \cosh m_\mathrm{in} R} \left(\frac{-B}{2} + \frac{Q^2}{64 \pi^2 M_\gamma R} \right), \\
  B &= (\phi_\mathrm{out} - \phi_\mathrm{in}) R - \frac{Q^2}{64 \pi^2 M_\gamma R}.
  \label{monopole-definition-screened}
\end{align}
We have already used the limit $m_\mathrm{in} R \gg 1$ to simplify $A$ and $B$, but not $\phi$.~Doing so, we find
\begin{equation}
  \phi(r) = \begin{cases}
  \bar \phi_\mathrm{in} & r \leq R, \\
  \bar \phi_\mathrm{out} - \frac{B}{r} - \frac{Q^2}{64 \pi^2 M_\gamma r^2} & r > R .
  \end{cases}
\end{equation}
We see that continuity of $\phi$ has been preserved, but now there is a discontinuity in $\phi'$ at $r = R$.~That being said, we derived this result from expressions that were continuous at $R$.~It is only in the limit $m_\mathrm{in} R \gg 1$ that the kink develops.~Finally, we note that continuity of $\phi$ at $r = R$, combined with Eq.~\eqref{monopole-definition-screened} sets $\bar \phi_\mathrm{in} = 0$, reducing the field solution to
\begin{equation}
  \phi(r) = \begin{cases}
    0 & r \leq R,\\
    \bar \phi_\mathrm{out} - \frac{B}{r} - \frac{Q^2}{64 \pi^2 M_\gamma r^2} & r > R .
  \end{cases}
\end{equation}

The perturbation from the matter interaction is
\begin{equation}
  \delta H_m = \begin{cases}
  0 & r < R, \\
  \frac{m_e}{M} (\bar \phi_\mathrm{out} - \frac{B}{r} -  \frac{Q^2}{64 \pi^2 M_\gamma r^2}) & r \geq R.
  \end{cases}
\end{equation}
The shift to the first energy level from $\delta H_m$ is
\begin{equation}
  \delta E_1^m = \frac{m_e}{M} \bar \phi_\mathrm{out} \left( 1 - \frac{4}{3} \frac{R^3}{a_0^3} \right) - \frac{m_e B}{M a_0} \left(1 - 2 \frac{R^2}{a_0^2} \right)
  - \frac{m_e Q^2}{64 \pi^2 M_\gamma M a_0^2} \left( \frac{1}{4} - \frac{R}{2 a_0} \right).
\end{equation}
Note that these expressions are not exact, as we have expanded them in $R / a_0$ and kept only the leading two terms.~This expression is to be compared with Eq.~\eqref{energy-level-shifts}.~The first term is akin to a varying-$\alpha$ theory (see, for example, \cite{Barrow:2014vva}) and is therefore not considered in this work.~The second term shows that the finite proton radius shifts the corresponding term in Eq.~\eqref{energy-level-shifts} by the entirely insignificant amount $\frac{R^2}{a_0^2} \approx 10^{-10}$.~The third term results from our inclusion of the $r^{-2}$ term in the exterior solution of $\phi$.~Clearly the finite radius correction is again negligible.~However, it is still unclear whether the leading order term can be {neglected} as was done in the main text.~This is only justified if the following is true:
\begin{equation}
  B \gg \frac{Q^2}{M_\gamma M a_0}.
\end{equation}
Substituting our solution for $B$, given by Eq.~\eqref{monopole-definition-screened}, we have
\begin{equation}
  \frac{1}{R} \gg \frac{1}{a_0},
\end{equation}
which is certainly true and therefore our approach was justified.~The same argument holds true for the $n = 2$ energy level perturbations as well.

Now we consider the energy level perturbations stemming from $\delta H_\gamma$.
The perturbation to the Hamiltonian via the electromagnetic interaction follows from Eq.~\eqref{V-and-perturbation} as
\begin{equation}
  \delta H_\gamma(r) = \begin{cases}
  \delta H_\gamma(R) & r < R, \\
  \frac{q Q}{4 \pi M_\gamma} \left( \frac{- \bar \phi_\mathrm{out}}{r} + \frac{B }{2 r^2} + \frac{Q^2}{192 \pi^2 M_\gamma r^3} \right) & r \geq R,
  \end{cases}
  \label{hamiltonian-pert-full}
\end{equation}
where $\delta H_\gamma(R)$ the exterior solution evaluated at $r = R$.~Once again we will expand the integrals for the shift to the energy levels in $R/a_0$, keeping only the leading terms.
The contribution from the interior part of the integrals, from 0 to $R$, scales as $(R / a_0)^3$, so this is once again strongly suppressed.~The same is true of the $1/r$ and $1/r^2$ pieces in the exterior solution, confirming again that we were justified in ignoring the finite size of the nucleus.

What is less clear is whether we were justified in dropping the $1/r^3$ term in Eq.~\eqref{hamiltonian-pert-full}.~The integrands for the energy perturbations diverge in the limit $R \to 0$, so we cannot easily Taylor expand in this limit as we did for the other terms.~Instead, we can note that the divergence is logarithmic, so the numerical coefficient after integration will be $\approx \log(R / a)$, so that the perturbation is
\begin{equation}
\label{eq:term1}
  \delta E \approx O(1) \times \frac{Q^4}{768 \pi^2 M_\gamma^2 a_0^3}.
\end{equation}
This is to be compared with the $B/r^2$ term in $\delta H_\gamma$, which ends up scaling as
\begin{equation}
\label{eq:term2}
  \delta E \approx \frac{Q^4}{M_\gamma^2 a_0^2 R}.
\end{equation}
Clearly {the term in Eq.~\eqref{eq:term2}} is much larger than {the term in Eq.~\eqref{eq:term1}}, so we confirm that it is not necessary to include the $1/r^3$ exterior potential. 

\subsection*{Unscreened case}
When the proton is unscreened, the full solution for the scalar field is
\begin{equation}
  \phi(r) = \begin{cases}
  \bar \phi_\mathrm{out} + \frac{3 Q^2}{128 \pi^2 R^2 M_\gamma } + \frac{J}{6} (r^2 - 3 R^2) - \frac{Q^2 r^4}{640 \pi^2 M_\gamma R^6} & r \leq R, \\
  \bar \phi_\mathrm{out} - \frac{B}{r} - \frac{Q^2}{64 \pi^2 M_\gamma r^2} & r > R ,
  \end{cases}
\end{equation}
where
\begin{equation}
  B = \frac{J R^3}{3} - \frac{3 Q^2}{80 \pi^2 M_\gamma R}.
\end{equation}
Unlike the screened case, where we assumed $m_\mathrm{in} R \gg 1$ in order to simplify the expression, there is no need to do so in this case.~This is because we have already neglected the scalar mass:~when we expanded the Lagrangian, we used $V_\mathrm{eff}(\phi) = V(\bar \phi) + J \varphi + \frac{1}{2} m_\mathrm{eff}^2 \varphi^2$, and we {neglected} $m_\mathrm{eff}$ as it is not the leading term.

Most of the arguments for the screened case apply here as well:~we are generally not interested in effects following from $\bar \phi_\mathrm{out}$ by itself as this is tantamount to a varying-$\alpha$ theory, and the $1/r^2$ term in the exterior solution may be safely neglected.~The only new effects for this case come from the terms in the interior solution.~However, these contribute corrections that are suppressed by at least $R / a_0$ relative to other terms, and therefore finite radius corrections may be neglected in the unscreened case as well.

\renewcommand{\em}{}
\addcontentsline{toc}{section}{References}
\bibliography{main}

\begin{thebibliography}{38}%
\makeatletter
\providecommand \@ifxundefined [1]{%
 \@ifx{#1\undefined}
}%
\providecommand \@ifnum [1]{%
 \ifnum #1\expandafter \@firstoftwo
 \else \expandafter \@secondoftwo
 \fi
}%
\providecommand \@ifx [1]{%
 \ifx #1\expandafter \@firstoftwo
 \else \expandafter \@secondoftwo
 \fi
}%
\providecommand \natexlab [1]{#1}%
\providecommand \enquote  [1]{``#1''}%
\providecommand \bibnamefont  [1]{#1}%
\providecommand \bibfnamefont [1]{#1}%
\providecommand \citenamefont [1]{#1}%
\providecommand \href@noop [0]{\@secondoftwo}%
\providecommand \href [0]{\begingroup \@sanitize@url \@href}%
\providecommand \@href[1]{\@@startlink{#1}\@@href}%
\providecommand \@@href[1]{\endgroup#1\@@endlink}%
\providecommand \@sanitize@url [0]{\catcode `\\12\catcode `\$12\catcode
  `\&12\catcode `\#12\catcode `\^12\catcode `\_12\catcode `\%12\relax}%
\providecommand \@@startlink[1]{}%
\providecommand \@@endlink[0]{}%
\providecommand \url  [0]{\begingroup\@sanitize@url \@url }%
\providecommand \@url [1]{\endgroup\@href {#1}{\urlprefix }}%
\providecommand \urlprefix  [0]{URL }%
\providecommand \Eprint [0]{\href }%
\providecommand \doibase [0]{http://dx.doi.org/}%
\providecommand \selectlanguage [0]{\@gobble}%
\providecommand \bibinfo  [0]{\@secondoftwo}%
\providecommand \bibfield  [0]{\@secondoftwo}%
\providecommand \translation [1]{[#1]}%
\providecommand \BibitemOpen [0]{}%
\providecommand \bibitemStop [0]{}%
\providecommand \bibitemNoStop [0]{.\EOS\space}%
\providecommand \EOS [0]{\spacefactor3000\relax}%
\providecommand \BibitemShut  [1]{\csname bibitem#1\endcsname}%
\let\auto@bib@innerbib\@empty
\bibitem [{\citenamefont {Adelberger}\ \emph {et~al.}(2003)\citenamefont
  {Adelberger}, \citenamefont {Heckel},\ and\ \citenamefont
  {Nelson}}]{Adelberger:2003zx}%
  \BibitemOpen
  \bibfield  {author} {\bibinfo {author} {\bibfnamefont {E.~G.}\ \bibnamefont
  {Adelberger}}, \bibinfo {author} {\bibfnamefont {B.~R.}\ \bibnamefont
  {Heckel}}, \ and\ \bibinfo {author} {\bibfnamefont {A.~E.}\ \bibnamefont
  {Nelson}},\ }\href {\doibase 10.1146/annurev.nucl.53.041002.110503}
  {\bibfield  {journal} {\bibinfo  {journal} {Ann. Rev. Nucl. Part. Sci.}\
  }\textbf {\bibinfo {volume} {53}},\ \bibinfo {pages} {77} (\bibinfo {year}
  {2003})},\ \Eprint {http://arxiv.org/abs/hep-ph/0307284}
  {arXiv:hep-ph/0307284} \BibitemShut {NoStop}%
\bibitem [{\citenamefont {Kapner}\ \emph {et~al.}(2007)\citenamefont {Kapner},
  \citenamefont {Cook}, \citenamefont {Adelberger}, \citenamefont {Gundlach},
  \citenamefont {Heckel}, \citenamefont {Hoyle},\ and\ \citenamefont
  {Swanson}}]{Kapner:2006si}%
  \BibitemOpen
  \bibfield  {author} {\bibinfo {author} {\bibfnamefont {D.~J.}\ \bibnamefont
  {Kapner}}, \bibinfo {author} {\bibfnamefont {T.~S.}\ \bibnamefont {Cook}},
  \bibinfo {author} {\bibfnamefont {E.~G.}\ \bibnamefont {Adelberger}},
  \bibinfo {author} {\bibfnamefont {J.~H.}\ \bibnamefont {Gundlach}}, \bibinfo
  {author} {\bibfnamefont {B.~R.}\ \bibnamefont {Heckel}}, \bibinfo {author}
  {\bibfnamefont {C.~D.}\ \bibnamefont {Hoyle}}, \ and\ \bibinfo {author}
  {\bibfnamefont {H.~E.}\ \bibnamefont {Swanson}},\ }\href {\doibase
  10.1103/PhysRevLett.98.021101} {\bibfield  {journal} {\bibinfo  {journal}
  {Phys. Rev. Lett.}\ }\textbf {\bibinfo {volume} {98}},\ \bibinfo {pages}
  {021101} (\bibinfo {year} {2007})},\ \Eprint
  {http://arxiv.org/abs/hep-ph/0611184} {arXiv:hep-ph/0611184} \BibitemShut
  {NoStop}%
\bibitem [{\citenamefont {Saridakis}\ \emph {et~al.}(2021)\citenamefont
  {Saridakis} \emph {et~al.}}]{CANTATA:2021ktz}%
  \BibitemOpen
  \bibfield  {author} {\bibinfo {author} {\bibfnamefont {E.~N.}\ \bibnamefont
  {Saridakis}} \emph {et~al.} (\bibinfo {collaboration} {CANTATA}),\
  }\href@noop {} {\  (\bibinfo {year} {2021})},\ \Eprint
  {http://arxiv.org/abs/2105.12582} {arXiv:2105.12582 [gr-qc]} \BibitemShut
  {NoStop}%
\bibitem [{\citenamefont {Joyce}\ \emph {et~al.}(2015)\citenamefont {Joyce},
  \citenamefont {Jain}, \citenamefont {Khoury},\ and\ \citenamefont
  {Trodden}}]{Joyce:2014kja}%
  \BibitemOpen
  \bibfield  {author} {\bibinfo {author} {\bibfnamefont {A.}~\bibnamefont
  {Joyce}}, \bibinfo {author} {\bibfnamefont {B.}~\bibnamefont {Jain}},
  \bibinfo {author} {\bibfnamefont {J.}~\bibnamefont {Khoury}}, \ and\ \bibinfo
  {author} {\bibfnamefont {M.}~\bibnamefont {Trodden}},\ }\href {\doibase
  10.1016/j.physrep.2014.12.002} {\bibfield  {journal} {\bibinfo  {journal}
  {Phys. Rept.}\ }\textbf {\bibinfo {volume} {568}},\ \bibinfo {pages} {1}
  (\bibinfo {year} {2015})},\ \Eprint {http://arxiv.org/abs/1407.0059}
  {arXiv:1407.0059 [astro-ph.CO]} \BibitemShut {NoStop}%
\bibitem [{\citenamefont {Khoury}\ and\ \citenamefont
  {Weltman}(2004)}]{Khoury:2003aq}%
  \BibitemOpen
  \bibfield  {author} {\bibinfo {author} {\bibfnamefont {J.}~\bibnamefont
  {Khoury}}\ and\ \bibinfo {author} {\bibfnamefont {A.}~\bibnamefont
  {Weltman}},\ }\href {\doibase 10.1103/PhysRevLett.93.171104} {\bibfield
  {journal} {\bibinfo  {journal} {Phys. Rev. Lett.}\ }\textbf {\bibinfo
  {volume} {93}},\ \bibinfo {pages} {171104} (\bibinfo {year} {2004})},\
  \Eprint {http://arxiv.org/abs/astro-ph/0309300} {arXiv:astro-ph/0309300}
  \BibitemShut {NoStop}%
\bibitem [{\citenamefont {Hinterbichler}\ and\ \citenamefont
  {Khoury}(2010)}]{Hinterbichler:2010es}%
  \BibitemOpen
  \bibfield  {author} {\bibinfo {author} {\bibfnamefont {K.}~\bibnamefont
  {Hinterbichler}}\ and\ \bibinfo {author} {\bibfnamefont {J.}~\bibnamefont
  {Khoury}},\ }\href {\doibase 10.1103/PhysRevLett.104.231301} {\bibfield
  {journal} {\bibinfo  {journal} {Phys. Rev. Lett.}\ }\textbf {\bibinfo
  {volume} {104}},\ \bibinfo {pages} {231301} (\bibinfo {year} {2010})},\
  \Eprint {http://arxiv.org/abs/1001.4525} {arXiv:1001.4525 [hep-th]}
  \BibitemShut {NoStop}%
\bibitem [{\citenamefont {Nicolis}\ \emph {et~al.}(2009)\citenamefont
  {Nicolis}, \citenamefont {Rattazzi},\ and\ \citenamefont
  {Trincherini}}]{Nicolis:2008in}%
  \BibitemOpen
  \bibfield  {author} {\bibinfo {author} {\bibfnamefont {A.}~\bibnamefont
  {Nicolis}}, \bibinfo {author} {\bibfnamefont {R.}~\bibnamefont {Rattazzi}}, \
  and\ \bibinfo {author} {\bibfnamefont {E.}~\bibnamefont {Trincherini}},\
  }\href {\doibase 10.1103/PhysRevD.79.064036} {\bibfield  {journal} {\bibinfo
  {journal} {Phys. Rev. D}\ }\textbf {\bibinfo {volume} {79}},\ \bibinfo
  {pages} {064036} (\bibinfo {year} {2009})},\ \Eprint
  {http://arxiv.org/abs/0811.2197} {arXiv:0811.2197 [hep-th]} \BibitemShut
  {NoStop}%
\bibitem [{\citenamefont {Burrage}\ and\ \citenamefont
  {Sakstein}(2016)}]{Burrage:2016bwy}%
  \BibitemOpen
  \bibfield  {author} {\bibinfo {author} {\bibfnamefont {C.}~\bibnamefont
  {Burrage}}\ and\ \bibinfo {author} {\bibfnamefont {J.}~\bibnamefont
  {Sakstein}},\ }\href {\doibase 10.1088/1475-7516/2016/11/045} {\bibfield
  {journal} {\bibinfo  {journal} {J. Cosmol. Astropart. Phys.}\ }\textbf
  {\bibinfo {volume} {11}},\ \bibinfo {pages} {045} (\bibinfo {year} {2016})},\
  \Eprint {http://arxiv.org/abs/1609.01192} {arXiv:1609.01192} \BibitemShut
  {NoStop}%
\bibitem [{\citenamefont {Burrage}\ and\ \citenamefont
  {Sakstein}(2018)}]{Burrage:2017qrf}%
  \BibitemOpen
  \bibfield  {author} {\bibinfo {author} {\bibfnamefont {C.}~\bibnamefont
  {Burrage}}\ and\ \bibinfo {author} {\bibfnamefont {J.}~\bibnamefont
  {Sakstein}},\ }\href {\doibase 10.1007/s41114-018-0011-x} {\bibfield
  {journal} {\bibinfo  {journal} {Living Rev. Rel.}\ }\textbf {\bibinfo
  {volume} {21}},\ \bibinfo {pages} {1} (\bibinfo {year} {2018})},\ \Eprint
  {http://arxiv.org/abs/1709.09071} {arXiv:1709.09071 [astro-ph.CO]}
  \BibitemShut {NoStop}%
\bibitem [{\citenamefont {Sakstein}(2018)}]{Sakstein:2018fwz}%
  \BibitemOpen
  \bibfield  {author} {\bibinfo {author} {\bibfnamefont {J.}~\bibnamefont
  {Sakstein}},\ }\href {\doibase 10.1142/S0218271818480085} {\bibfield
  {journal} {\bibinfo  {journal} {Int. J. Mod. Phys. D}\ }\textbf {\bibinfo
  {volume} {27}},\ \bibinfo {pages} {1848008} (\bibinfo {year} {2018})},\
  \Eprint {http://arxiv.org/abs/2002.04194} {arXiv:2002.04194 [astro-ph.CO]}
  \BibitemShut {NoStop}%
\bibitem [{\citenamefont {Baker}\ \emph {et~al.}(2021)\citenamefont {Baker}
  \emph {et~al.}}]{Baker:2019gxo}%
  \BibitemOpen
  \bibfield  {author} {\bibinfo {author} {\bibfnamefont {T.}~\bibnamefont
  {Baker}} \emph {et~al.},\ }\href {\doibase 10.1103/RevModPhys.93.015003}
  {\bibfield  {journal} {\bibinfo  {journal} {Rev. Mod. Phys.}\ }\textbf
  {\bibinfo {volume} {93}},\ \bibinfo {pages} {015003} (\bibinfo {year}
  {2021})},\ \Eprint {http://arxiv.org/abs/1908.03430} {arXiv:1908.03430
  [astro-ph.CO]} \BibitemShut {NoStop}%
\bibitem [{\citenamefont {Brax}\ \emph {et~al.}(2021)\citenamefont {Brax},
  \citenamefont {Casas}, \citenamefont {Desmond},\ and\ \citenamefont
  {Elder}}]{Brax:2021wcv}%
  \BibitemOpen
  \bibfield  {author} {\bibinfo {author} {\bibfnamefont {P.}~\bibnamefont
  {Brax}}, \bibinfo {author} {\bibfnamefont {S.}~\bibnamefont {Casas}},
  \bibinfo {author} {\bibfnamefont {H.}~\bibnamefont {Desmond}}, \ and\
  \bibinfo {author} {\bibfnamefont {B.}~\bibnamefont {Elder}},\ }\href
  {\doibase 10.3390/universe8010011} {\bibfield  {journal} {\bibinfo  {journal}
  {Universe}\ }\textbf {\bibinfo {volume} {8}},\ \bibinfo {pages} {11}
  (\bibinfo {year} {2021})},\ \Eprint {http://arxiv.org/abs/2201.10817}
  {arXiv:2201.10817 [gr-qc]} \BibitemShut {NoStop}%
\bibitem [{\citenamefont {Vagnozzi}\ \emph {et~al.}(2021)\citenamefont
  {Vagnozzi}, \citenamefont {Visinelli}, \citenamefont {Brax}, \citenamefont
  {Davis},\ and\ \citenamefont {Sakstein}}]{Vagnozzi:2021quy}%
  \BibitemOpen
  \bibfield  {author} {\bibinfo {author} {\bibfnamefont {S.}~\bibnamefont
  {Vagnozzi}}, \bibinfo {author} {\bibfnamefont {L.}~\bibnamefont {Visinelli}},
  \bibinfo {author} {\bibfnamefont {P.}~\bibnamefont {Brax}}, \bibinfo {author}
  {\bibfnamefont {A.-C.}\ \bibnamefont {Davis}}, \ and\ \bibinfo {author}
  {\bibfnamefont {J.}~\bibnamefont {Sakstein}},\ }\href {\doibase
  10.1103/PhysRevD.104.063023} {\bibfield  {journal} {\bibinfo  {journal}
  {Phys. Rev. D}\ }\textbf {\bibinfo {volume} {104}},\ \bibinfo {pages}
  {063023} (\bibinfo {year} {2021})},\ \Eprint
  {http://arxiv.org/abs/2103.15834} {arXiv:2103.15834 [hep-ph]} \BibitemShut
  {NoStop}%
\bibitem [{\citenamefont {Anastassopoulos}\ \emph {et~al.}(2015)\citenamefont
  {Anastassopoulos} \emph {et~al.}}]{CAST:2015npk}%
  \BibitemOpen
  \bibfield  {author} {\bibinfo {author} {\bibfnamefont {V.}~\bibnamefont
  {Anastassopoulos}} \emph {et~al.} (\bibinfo {collaboration} {CAST}),\ }\href
  {\doibase 10.1016/j.physletb.2015.07.049} {\bibfield  {journal} {\bibinfo
  {journal} {Phys. Lett. B}\ }\textbf {\bibinfo {volume} {749}},\ \bibinfo
  {pages} {172} (\bibinfo {year} {2015})},\ \Eprint
  {http://arxiv.org/abs/1503.04561} {arXiv:1503.04561 [astro-ph.SR]}
  \BibitemShut {NoStop}%
\bibitem [{\citenamefont {Steffen}\ \emph {et~al.}(2010)\citenamefont
  {Steffen}, \citenamefont {Upadhye}, \citenamefont {Baumbaugh}, \citenamefont
  {Chou}, \citenamefont {Mazur}, \citenamefont {Tomlin}, \citenamefont
  {Weltman},\ and\ \citenamefont {Wester}}]{Steffen:2010ze}%
  \BibitemOpen
  \bibfield  {author} {\bibinfo {author} {\bibfnamefont {J.~H.}\ \bibnamefont
  {Steffen}}, \bibinfo {author} {\bibfnamefont {A.}~\bibnamefont {Upadhye}},
  \bibinfo {author} {\bibfnamefont {A.}~\bibnamefont {Baumbaugh}}, \bibinfo
  {author} {\bibfnamefont {A.~S.}\ \bibnamefont {Chou}}, \bibinfo {author}
  {\bibfnamefont {P.~O.}\ \bibnamefont {Mazur}}, \bibinfo {author}
  {\bibfnamefont {R.}~\bibnamefont {Tomlin}}, \bibinfo {author} {\bibfnamefont
  {A.}~\bibnamefont {Weltman}}, \ and\ \bibinfo {author} {\bibfnamefont
  {W.}~\bibnamefont {Wester}} (\bibinfo {collaboration} {GammeV}),\ }\href
  {\doibase 10.1103/PhysRevLett.105.261803} {\bibfield  {journal} {\bibinfo
  {journal} {Phys. Rev. Lett.}\ }\textbf {\bibinfo {volume} {105}},\ \bibinfo
  {pages} {261803} (\bibinfo {year} {2010})},\ \Eprint
  {http://arxiv.org/abs/1010.0988} {arXiv:1010.0988 [astro-ph.CO]} \BibitemShut
  {NoStop}%
\bibitem [{\citenamefont {Brax}\ \emph {et~al.}(2018)\citenamefont {Brax},
  \citenamefont {Davis}, \citenamefont {Elder},\ and\ \citenamefont
  {Wong}}]{Brax:2018zfb}%
  \BibitemOpen
  \bibfield  {author} {\bibinfo {author} {\bibfnamefont {P.}~\bibnamefont
  {Brax}}, \bibinfo {author} {\bibfnamefont {A.-C.}\ \bibnamefont {Davis}},
  \bibinfo {author} {\bibfnamefont {B.}~\bibnamefont {Elder}}, \ and\ \bibinfo
  {author} {\bibfnamefont {L.~K.}\ \bibnamefont {Wong}},\ }\href {\doibase
  10.1103/PhysRevD.97.084050} {\bibfield  {journal} {\bibinfo  {journal} {Phys.
  Rev. D}\ }\textbf {\bibinfo {volume} {97}},\ \bibinfo {pages} {084050}
  (\bibinfo {year} {2018})},\ \Eprint {http://arxiv.org/abs/1802.05545}
  {arXiv:1802.05545 [hep-ph]} \BibitemShut {NoStop}%
\bibitem [{\citenamefont {Zavattini}\ \emph {et~al.}(2006)\citenamefont
  {Zavattini} \emph {et~al.}}]{PVLAS:2005sku}%
  \BibitemOpen
  \bibfield  {author} {\bibinfo {author} {\bibfnamefont {E.}~\bibnamefont
  {Zavattini}} \emph {et~al.} (\bibinfo {collaboration} {PVLAS}),\ }\href
  {\doibase 10.1103/PhysRevLett.99.129901} {\bibfield  {journal} {\bibinfo
  {journal} {Phys. Rev. Lett.}\ }\textbf {\bibinfo {volume} {96}},\ \bibinfo
  {pages} {110406} (\bibinfo {year} {2006})},\ \bibinfo {note} {[Erratum:
  Phys.Rev.Lett. 99, 129901 (2007)]},\ \Eprint
  {http://arxiv.org/abs/hep-ex/0507107} {arXiv:hep-ex/0507107} \BibitemShut
  {NoStop}%
\bibitem [{\citenamefont {Brax}\ \emph
  {et~al.}(2007{\natexlab{a}})\citenamefont {Brax}, \citenamefont {van~de
  Bruck},\ and\ \citenamefont {Davis}}]{Brax:2007ak}%
  \BibitemOpen
  \bibfield  {author} {\bibinfo {author} {\bibfnamefont {P.}~\bibnamefont
  {Brax}}, \bibinfo {author} {\bibfnamefont {C.}~\bibnamefont {van~de Bruck}},
  \ and\ \bibinfo {author} {\bibfnamefont {A.-C.}\ \bibnamefont {Davis}},\
  }\href {\doibase 10.1103/PhysRevLett.99.121103} {\bibfield  {journal}
  {\bibinfo  {journal} {Phys. Rev. Lett.}\ }\textbf {\bibinfo {volume} {99}},\
  \bibinfo {pages} {121103} (\bibinfo {year} {2007}{\natexlab{a}})},\ \Eprint
  {http://arxiv.org/abs/hep-ph/0703243} {arXiv:hep-ph/0703243} \BibitemShut
  {NoStop}%
\bibitem [{\citenamefont {Brax}\ \emph
  {et~al.}(2007{\natexlab{b}})\citenamefont {Brax}, \citenamefont {van~de
  Bruck}, \citenamefont {Davis}, \citenamefont {Mota},\ and\ \citenamefont
  {Shaw}}]{Brax:2007hi}%
  \BibitemOpen
  \bibfield  {author} {\bibinfo {author} {\bibfnamefont {P.}~\bibnamefont
  {Brax}}, \bibinfo {author} {\bibfnamefont {C.}~\bibnamefont {van~de Bruck}},
  \bibinfo {author} {\bibfnamefont {A.-C.}\ \bibnamefont {Davis}}, \bibinfo
  {author} {\bibfnamefont {D.~F.}\ \bibnamefont {Mota}}, \ and\ \bibinfo
  {author} {\bibfnamefont {D.~J.}\ \bibnamefont {Shaw}},\ }\href {\doibase
  10.1103/PhysRevD.76.085010} {\bibfield  {journal} {\bibinfo  {journal} {Phys.
  Rev. D}\ }\textbf {\bibinfo {volume} {76}},\ \bibinfo {pages} {085010}
  (\bibinfo {year} {2007}{\natexlab{b}})},\ \Eprint
  {http://arxiv.org/abs/0707.2801} {arXiv:0707.2801 [hep-ph]} \BibitemShut
  {NoStop}%
\bibitem [{\citenamefont {Jaffe}\ \emph {et~al.}(2017)\citenamefont {Jaffe},
  \citenamefont {Haslinger}, \citenamefont {Xu}, \citenamefont {Hamilton},
  \citenamefont {Upadhye}, \citenamefont {Elder}, \citenamefont {Khoury},\ and\
  \citenamefont {Müller}}]{Jaffe:2016fsh}%
  \BibitemOpen
  \bibfield  {author} {\bibinfo {author} {\bibfnamefont {M.}~\bibnamefont
  {Jaffe}}, \bibinfo {author} {\bibfnamefont {P.}~\bibnamefont {Haslinger}},
  \bibinfo {author} {\bibfnamefont {V.}~\bibnamefont {Xu}}, \bibinfo {author}
  {\bibfnamefont {P.}~\bibnamefont {Hamilton}}, \bibinfo {author}
  {\bibfnamefont {A.}~\bibnamefont {Upadhye}}, \bibinfo {author} {\bibfnamefont
  {B.}~\bibnamefont {Elder}}, \bibinfo {author} {\bibfnamefont
  {J.}~\bibnamefont {Khoury}}, \ and\ \bibinfo {author} {\bibfnamefont
  {H.}~\bibnamefont {Müller}},\ }\href {\doibase 10.1038/nphys4189} {\bibfield
   {journal} {\bibinfo  {journal} {Nat. Phys.}\ }\textbf {\bibinfo {volume}
  {13}},\ \bibinfo {pages} {938} (\bibinfo {year} {2017})},\ \Eprint
  {http://arxiv.org/abs/1612.05171} {arXiv:1612.05171} \BibitemShut {NoStop}%
\bibitem [{\citenamefont {Sabulsky}\ \emph {et~al.}(2019)\citenamefont
  {Sabulsky}, \citenamefont {Dutta}, \citenamefont {Hinds}, \citenamefont
  {Elder}, \citenamefont {Burrage},\ and\ \citenamefont
  {Copeland}}]{Sabulsky:2018jma}%
  \BibitemOpen
  \bibfield  {author} {\bibinfo {author} {\bibfnamefont {D.~O.}\ \bibnamefont
  {Sabulsky}}, \bibinfo {author} {\bibfnamefont {I.}~\bibnamefont {Dutta}},
  \bibinfo {author} {\bibfnamefont {E.~A.}\ \bibnamefont {Hinds}}, \bibinfo
  {author} {\bibfnamefont {B.}~\bibnamefont {Elder}}, \bibinfo {author}
  {\bibfnamefont {C.}~\bibnamefont {Burrage}}, \ and\ \bibinfo {author}
  {\bibfnamefont {E.~J.}\ \bibnamefont {Copeland}},\ }\href {\doibase
  10.1103/PhysRevLett.123.061102} {\bibfield  {journal} {\bibinfo  {journal}
  {Phys. Rev. Lett.}\ }\textbf {\bibinfo {volume} {123}},\ \bibinfo {pages}
  {061102} (\bibinfo {year} {2019})},\ \Eprint
  {http://arxiv.org/abs/1812.08244} {arXiv:1812.08244 [physics.atom-ph]}
  \BibitemShut {NoStop}%
\bibitem [{\citenamefont {{Parthey}}\ \emph {et~al.}(2011)\citenamefont
  {{Parthey}}, \citenamefont {{Matveev}}, \citenamefont {{Alnis}},
  \citenamefont {{Bernhardt}}, \citenamefont {{Beyer}}, \citenamefont
  {{Holzwarth}}, \citenamefont {{Maistrou}}, \citenamefont {{Pohl}},
  \citenamefont {{Predehl}}, \citenamefont {{Udem}}, \citenamefont {{Wilken}},
  \citenamefont {{Kolachevsky}}, \citenamefont {{Abgrall}}, \citenamefont
  {{Rovera}}, \citenamefont {{Salomon}}, \citenamefont {{Laurent}},\ and\
  \citenamefont {{H{\"a}nsch}}}]{Parthey_2011PhRvL.107t3001P}%
  \BibitemOpen
  \bibfield  {author} {\bibinfo {author} {\bibfnamefont {C.~G.}\ \bibnamefont
  {{Parthey}}}, \bibinfo {author} {\bibfnamefont {A.}~\bibnamefont
  {{Matveev}}}, \bibinfo {author} {\bibfnamefont {J.}~\bibnamefont {{Alnis}}},
  \bibinfo {author} {\bibfnamefont {B.}~\bibnamefont {{Bernhardt}}}, \bibinfo
  {author} {\bibfnamefont {A.}~\bibnamefont {{Beyer}}}, \bibinfo {author}
  {\bibfnamefont {R.}~\bibnamefont {{Holzwarth}}}, \bibinfo {author}
  {\bibfnamefont {A.}~\bibnamefont {{Maistrou}}}, \bibinfo {author}
  {\bibfnamefont {R.}~\bibnamefont {{Pohl}}}, \bibinfo {author} {\bibfnamefont
  {K.}~\bibnamefont {{Predehl}}}, \bibinfo {author} {\bibfnamefont
  {T.}~\bibnamefont {{Udem}}}, \bibinfo {author} {\bibfnamefont
  {T.}~\bibnamefont {{Wilken}}}, \bibinfo {author} {\bibfnamefont
  {N.}~\bibnamefont {{Kolachevsky}}}, \bibinfo {author} {\bibfnamefont
  {M.}~\bibnamefont {{Abgrall}}}, \bibinfo {author} {\bibfnamefont
  {D.}~\bibnamefont {{Rovera}}}, \bibinfo {author} {\bibfnamefont
  {C.}~\bibnamefont {{Salomon}}}, \bibinfo {author} {\bibfnamefont
  {P.}~\bibnamefont {{Laurent}}}, \ and\ \bibinfo {author} {\bibfnamefont
  {T.~W.}\ \bibnamefont {{H{\"a}nsch}}},\ }\href {\doibase
  10.1103/PhysRevLett.107.203001} {\bibfield  {journal} {\bibinfo  {journal}
  {\prl}\ }\textbf {\bibinfo {volume} {107}},\ \bibinfo {eid} {203001}
  (\bibinfo {year} {2011})},\ \Eprint {http://arxiv.org/abs/1107.3101}
  {arXiv:1107.3101 [physics.atom-ph]} \BibitemShut {NoStop}%
\bibitem [{\citenamefont {Brax}\ and\ \citenamefont
  {Burrage}(2011)}]{Brax:2010gp}%
  \BibitemOpen
  \bibfield  {author} {\bibinfo {author} {\bibfnamefont {P.}~\bibnamefont
  {Brax}}\ and\ \bibinfo {author} {\bibfnamefont {C.}~\bibnamefont {Burrage}},\
  }\href {\doibase 10.1103/PhysRevD.83.035020} {\bibfield  {journal} {\bibinfo
  {journal} {Phys. Rev. D}\ }\textbf {\bibinfo {volume} {83}},\ \bibinfo
  {pages} {035020} (\bibinfo {year} {2011})},\ \Eprint
  {http://arxiv.org/abs/1010.5108} {arXiv:1010.5108} \BibitemShut {NoStop}%
\bibitem [{\citenamefont {Brax}\ \emph {et~al.}(2022)\citenamefont {Brax},
  \citenamefont {Davis},\ and\ \citenamefont {Elder}}]{Brax:2022olf}%
  \BibitemOpen
  \bibfield  {author} {\bibinfo {author} {\bibfnamefont {P.}~\bibnamefont
  {Brax}}, \bibinfo {author} {\bibfnamefont {A.-C.}\ \bibnamefont {Davis}}, \
  and\ \bibinfo {author} {\bibfnamefont {B.}~\bibnamefont {Elder}},\
  }\href@noop {} {\  (\bibinfo {year} {2022})},\ \Eprint
  {http://arxiv.org/abs/2207.11633} {arXiv:2207.11633 [hep-ph]} \BibitemShut
  {NoStop}%
\bibitem [{\citenamefont {Wong}\ and\ \citenamefont
  {Davis}(2017)}]{Wong:2017jer}%
  \BibitemOpen
  \bibfield  {author} {\bibinfo {author} {\bibfnamefont {L.~K.}\ \bibnamefont
  {Wong}}\ and\ \bibinfo {author} {\bibfnamefont {A.-C.}\ \bibnamefont
  {Davis}},\ }\href {\doibase 10.1103/PhysRevD.95.104010} {\bibfield  {journal}
  {\bibinfo  {journal} {Phys. Rev. D}\ }\textbf {\bibinfo {volume} {95}},\
  \bibinfo {pages} {104010} (\bibinfo {year} {2017})},\ \Eprint
  {http://arxiv.org/abs/1703.05659} {arXiv:1703.05659 [astro-ph.CO]}
  \BibitemShut {NoStop}%
\bibitem [{\citenamefont {Hamilton}\ \emph {et~al.}(2015)\citenamefont
  {Hamilton}, \citenamefont {Jaffe}, \citenamefont {Haslinger}, \citenamefont
  {Simmons}, \citenamefont {M\"uller},\ and\ \citenamefont
  {Khoury}}]{Hamilton:2015zga}%
  \BibitemOpen
  \bibfield  {author} {\bibinfo {author} {\bibfnamefont {P.}~\bibnamefont
  {Hamilton}}, \bibinfo {author} {\bibfnamefont {M.}~\bibnamefont {Jaffe}},
  \bibinfo {author} {\bibfnamefont {P.}~\bibnamefont {Haslinger}}, \bibinfo
  {author} {\bibfnamefont {Q.}~\bibnamefont {Simmons}}, \bibinfo {author}
  {\bibfnamefont {H.}~\bibnamefont {M\"uller}}, \ and\ \bibinfo {author}
  {\bibfnamefont {J.}~\bibnamefont {Khoury}},\ }\href {\doibase
  10.1126/science.aaa8883} {\bibfield  {journal} {\bibinfo  {journal}
  {Science}\ }\textbf {\bibinfo {volume} {349}},\ \bibinfo {pages} {849}
  (\bibinfo {year} {2015})},\ \Eprint {http://arxiv.org/abs/1502.03888}
  {arXiv:1502.03888 [physics.atom-ph]} \BibitemShut {NoStop}%
\bibitem [{\citenamefont {Elder}\ \emph {et~al.}(2016)\citenamefont {Elder},
  \citenamefont {Khoury}, \citenamefont {Haslinger}, \citenamefont {Jaffe},
  \citenamefont {M\"uller},\ and\ \citenamefont {Hamilton}}]{Elder:2016yxm}%
  \BibitemOpen
  \bibfield  {author} {\bibinfo {author} {\bibfnamefont {B.}~\bibnamefont
  {Elder}}, \bibinfo {author} {\bibfnamefont {J.}~\bibnamefont {Khoury}},
  \bibinfo {author} {\bibfnamefont {P.}~\bibnamefont {Haslinger}}, \bibinfo
  {author} {\bibfnamefont {M.}~\bibnamefont {Jaffe}}, \bibinfo {author}
  {\bibfnamefont {H.}~\bibnamefont {M\"uller}}, \ and\ \bibinfo {author}
  {\bibfnamefont {P.}~\bibnamefont {Hamilton}},\ }\href {\doibase
  10.1103/PhysRevD.94.044051} {\bibfield  {journal} {\bibinfo  {journal} {Phys.
  Rev. D}\ }\textbf {\bibinfo {volume} {94}},\ \bibinfo {pages} {044051}
  (\bibinfo {year} {2016})},\ \Eprint {http://arxiv.org/abs/1603.06587}
  {arXiv:1603.06587 [astro-ph.CO]} \BibitemShut {NoStop}%
\bibitem [{\citenamefont {Parthey}\ \emph {et~al.}(2010)\citenamefont
  {Parthey}, \citenamefont {Matveev}, \citenamefont {Alnis}, \citenamefont
  {Pohl}, \citenamefont {Udem}, \citenamefont {Jentschura}, \citenamefont
  {Kolachevsky},\ and\ \citenamefont {H\"ansch}}]{Parthey:2010aya}%
  \BibitemOpen
  \bibfield  {author} {\bibinfo {author} {\bibfnamefont {C.~G.}\ \bibnamefont
  {Parthey}}, \bibinfo {author} {\bibfnamefont {A.}~\bibnamefont {Matveev}},
  \bibinfo {author} {\bibfnamefont {J.}~\bibnamefont {Alnis}}, \bibinfo
  {author} {\bibfnamefont {R.}~\bibnamefont {Pohl}}, \bibinfo {author}
  {\bibfnamefont {T.}~\bibnamefont {Udem}}, \bibinfo {author} {\bibfnamefont
  {U.~D.}\ \bibnamefont {Jentschura}}, \bibinfo {author} {\bibfnamefont
  {N.}~\bibnamefont {Kolachevsky}}, \ and\ \bibinfo {author} {\bibfnamefont
  {T.~W.}\ \bibnamefont {H\"ansch}},\ }\href {\doibase
  10.1103/PhysRevLett.104.233001} {\bibfield  {journal} {\bibinfo  {journal}
  {Phys. Rev. Lett.}\ }\textbf {\bibinfo {volume} {104}},\ \bibinfo {pages}
  {233001} (\bibinfo {year} {2010})}\BibitemShut {NoStop}%
\bibitem [{\citenamefont {Yin}\ \emph {et~al.}(2022)\citenamefont {Yin},
  \citenamefont {Li}, \citenamefont {Yin}, \citenamefont {Xu}, \citenamefont
  {Bian}, \citenamefont {Xie}, \citenamefont {Duan}, \citenamefont {Huang},
  \citenamefont {He},\ and\ \citenamefont {Du}}]{Yin:2022geb}%
  \BibitemOpen
  \bibfield  {author} {\bibinfo {author} {\bibfnamefont {P.}~\bibnamefont
  {Yin}}, \bibinfo {author} {\bibfnamefont {R.}~\bibnamefont {Li}}, \bibinfo
  {author} {\bibfnamefont {C.}~\bibnamefont {Yin}}, \bibinfo {author}
  {\bibfnamefont {X.}~\bibnamefont {Xu}}, \bibinfo {author} {\bibfnamefont
  {X.}~\bibnamefont {Bian}}, \bibinfo {author} {\bibfnamefont {H.}~\bibnamefont
  {Xie}}, \bibinfo {author} {\bibfnamefont {C.-K.}\ \bibnamefont {Duan}},
  \bibinfo {author} {\bibfnamefont {P.}~\bibnamefont {Huang}}, \bibinfo
  {author} {\bibfnamefont {J.-h.}\ \bibnamefont {He}}, \ and\ \bibinfo {author}
  {\bibfnamefont {J.}~\bibnamefont {Du}},\ }\href {\doibase
  10.1038/s41567-022-01706-9} {\bibfield  {journal} {\bibinfo  {journal}
  {Nature Phys.}\ }\textbf {\bibinfo {volume} {18}},\ \bibinfo {pages} {1181}
  (\bibinfo {year} {2022})}\BibitemShut {NoStop}%
\bibitem [{\citenamefont {Brax}\ \emph {et~al.}(2023)\citenamefont {Brax},
  \citenamefont {Davis},\ and\ \citenamefont {Elder}}]{Brax:2022uiv}%
  \BibitemOpen
  \bibfield  {author} {\bibinfo {author} {\bibfnamefont {P.}~\bibnamefont
  {Brax}}, \bibinfo {author} {\bibfnamefont {A.-C.}\ \bibnamefont {Davis}}, \
  and\ \bibinfo {author} {\bibfnamefont {B.}~\bibnamefont {Elder}},\ }\href
  {\doibase 10.1103/PhysRevD.107.084025} {\bibfield  {journal} {\bibinfo
  {journal} {Phys. Rev. D}\ }\textbf {\bibinfo {volume} {107}},\ \bibinfo
  {pages} {084025} (\bibinfo {year} {2023})},\ \Eprint
  {http://arxiv.org/abs/2211.07840} {arXiv:2211.07840 [gr-qc]} \BibitemShut
  {NoStop}%
\bibitem [{\citenamefont {Chou}\ \emph {et~al.}(2009)\citenamefont {Chou} \emph
  {et~al.}}]{GammeV:2008cqp}%
  \BibitemOpen
  \bibfield  {author} {\bibinfo {author} {\bibfnamefont {A.~S.}\ \bibnamefont
  {Chou}} \emph {et~al.} (\bibinfo {collaboration} {GammeV}),\ }\href {\doibase
  10.1103/PhysRevLett.102.030402} {\bibfield  {journal} {\bibinfo  {journal}
  {Phys. Rev. Lett.}\ }\textbf {\bibinfo {volume} {102}},\ \bibinfo {pages}
  {030402} (\bibinfo {year} {2009})},\ \Eprint {http://arxiv.org/abs/0806.2438}
  {arXiv:0806.2438 [hep-ex]} \BibitemShut {NoStop}%
\bibitem [{\citenamefont {Brax}\ \emph {et~al.}(2009)\citenamefont {Brax},
  \citenamefont {Burrage}, \citenamefont {Davis}, \citenamefont {Seery},\ and\
  \citenamefont {Weltman}}]{Brax:2009aw}%
  \BibitemOpen
  \bibfield  {author} {\bibinfo {author} {\bibfnamefont {P.}~\bibnamefont
  {Brax}}, \bibinfo {author} {\bibfnamefont {C.}~\bibnamefont {Burrage}},
  \bibinfo {author} {\bibfnamefont {A.-C.}\ \bibnamefont {Davis}}, \bibinfo
  {author} {\bibfnamefont {D.}~\bibnamefont {Seery}}, \ and\ \bibinfo {author}
  {\bibfnamefont {A.}~\bibnamefont {Weltman}},\ }\href
  {https://doi.org/10.1088/1126-6708/2009/09/128} {\bibfield  {journal}
  {\bibinfo  {journal} {J. High Energy Phys.}\ }\textbf {\bibinfo {volume}
  {09}},\ \bibinfo {pages} {128} (\bibinfo {year} {2009})},\ \Eprint
  {http://arxiv.org/abs/0904.3002} {arXiv:0904.3002} \BibitemShut {NoStop}%
\bibitem [{\citenamefont {{Hanneke}}\ \emph {et~al.}(2008)\citenamefont
  {{Hanneke}}, \citenamefont {{Fogwell}},\ and\ \citenamefont
  {{Gabrielse}}}]{2008PhRvL.100l0801H}%
  \BibitemOpen
  \bibfield  {author} {\bibinfo {author} {\bibfnamefont {D.}~\bibnamefont
  {{Hanneke}}}, \bibinfo {author} {\bibfnamefont {S.}~\bibnamefont
  {{Fogwell}}}, \ and\ \bibinfo {author} {\bibfnamefont {G.}~\bibnamefont
  {{Gabrielse}}},\ }\href {\doibase 10.1103/PhysRevLett.100.120801} {\bibfield
  {journal} {\bibinfo  {journal} {\prl}\ }\textbf {\bibinfo {volume} {100}},\
  \bibinfo {eid} {120801} (\bibinfo {year} {2008})},\ \Eprint
  {http://arxiv.org/abs/0801.1134} {arXiv:0801.1134 [physics.atom-ph]}
  \BibitemShut {NoStop}%
\bibitem [{\citenamefont {Hanneke}\ \emph {et~al.}(2011)\citenamefont
  {Hanneke}, \citenamefont {Hoogerheide},\ and\ \citenamefont
  {Gabrielse}}]{Hanneke:2010au}%
  \BibitemOpen
  \bibfield  {author} {\bibinfo {author} {\bibfnamefont {D.}~\bibnamefont
  {Hanneke}}, \bibinfo {author} {\bibfnamefont {S.~F.}\ \bibnamefont
  {Hoogerheide}}, \ and\ \bibinfo {author} {\bibfnamefont {G.}~\bibnamefont
  {Gabrielse}},\ }\href {\doibase 10.1103/PhysRevA.83.052122} {\bibfield
  {journal} {\bibinfo  {journal} {Phys. Rev. A}\ }\textbf {\bibinfo {volume}
  {83}},\ \bibinfo {pages} {052122} (\bibinfo {year} {2011})},\ \Eprint
  {http://arxiv.org/abs/1009.4831} {arXiv:1009.4831 [physics.atom-ph]}
  \BibitemShut {NoStop}%
\bibitem [{\citenamefont {Burrage}\ \emph
  {et~al.}(2009{\natexlab{a}})\citenamefont {Burrage}, \citenamefont {Davis},\
  and\ \citenamefont {Shaw}}]{Burrage:2008ii}%
  \BibitemOpen
  \bibfield  {author} {\bibinfo {author} {\bibfnamefont {C.}~\bibnamefont
  {Burrage}}, \bibinfo {author} {\bibfnamefont {A.-C.}\ \bibnamefont {Davis}},
  \ and\ \bibinfo {author} {\bibfnamefont {D.~J.}\ \bibnamefont {Shaw}},\
  }\href {\doibase 10.1103/PhysRevD.79.044028} {\bibfield  {journal} {\bibinfo
  {journal} {Phys. Rev. D}\ }\textbf {\bibinfo {volume} {79}},\ \bibinfo
  {pages} {044028} (\bibinfo {year} {2009}{\natexlab{a}})},\ \Eprint
  {http://arxiv.org/abs/0809.1763} {arXiv:0809.1763 [astro-ph]} \BibitemShut
  {NoStop}%
\bibitem [{\citenamefont {Burrage}\ \emph
  {et~al.}(2009{\natexlab{b}})\citenamefont {Burrage}, \citenamefont {Davis},\
  and\ \citenamefont {Shaw}}]{Burrage:2009mj}%
  \BibitemOpen
  \bibfield  {author} {\bibinfo {author} {\bibfnamefont {C.}~\bibnamefont
  {Burrage}}, \bibinfo {author} {\bibfnamefont {A.-C.}\ \bibnamefont {Davis}},
  \ and\ \bibinfo {author} {\bibfnamefont {D.~J.}\ \bibnamefont {Shaw}},\
  }\href {\doibase 10.1103/PhysRevLett.102.201101} {\bibfield  {journal}
  {\bibinfo  {journal} {Phys. Rev. Lett.}\ }\textbf {\bibinfo {volume} {102}},\
  \bibinfo {pages} {201101} (\bibinfo {year} {2009}{\natexlab{b}})},\ \Eprint
  {http://arxiv.org/abs/0902.2320} {arXiv:0902.2320 [astro-ph.CO]} \BibitemShut
  {NoStop}%
\bibitem [{\citenamefont {Pettinari}\ and\ \citenamefont
  {Crittenden}(2010)}]{Pettinari:2010ay}%
  \BibitemOpen
  \bibfield  {author} {\bibinfo {author} {\bibfnamefont {G.~W.}\ \bibnamefont
  {Pettinari}}\ and\ \bibinfo {author} {\bibfnamefont {R.}~\bibnamefont
  {Crittenden}},\ }\href {\doibase 10.1103/PhysRevD.82.083502} {\bibfield
  {journal} {\bibinfo  {journal} {Phys. Rev. D}\ }\textbf {\bibinfo {volume}
  {82}},\ \bibinfo {pages} {083502} (\bibinfo {year} {2010})},\ \Eprint
  {http://arxiv.org/abs/1007.0024} {arXiv:1007.0024 [astro-ph.CO]} \BibitemShut
  {NoStop}%
\bibitem [{\citenamefont {Barrow}\ and\ \citenamefont
  {Magueijo}(2015)}]{Barrow:2014vva}%
  \BibitemOpen
  \bibfield  {author} {\bibinfo {author} {\bibfnamefont {J.~D.}\ \bibnamefont
  {Barrow}}\ and\ \bibinfo {author} {\bibfnamefont {J.~a.}\ \bibnamefont
  {Magueijo}},\ }\href {\doibase 10.1142/S0217732315400295} {\bibfield
  {journal} {\bibinfo  {journal} {Mod. Phys. Lett. A}\ }\textbf {\bibinfo
  {volume} {30}},\ \bibinfo {pages} {1540029} (\bibinfo {year} {2015})},\
  \Eprint {http://arxiv.org/abs/1412.3278} {arXiv:1412.3278 [gr-qc]}
  \BibitemShut {NoStop}%
\end{thebibliography}%
\end{document}